\documentclass[aps,twocolumn,prl,showpacs,preprintnumbers,amsmath,amssymb,floatfix]{revtex4} 
\usepackage{graphicx}
\usepackage{dcolumn}
\usepackage{bm}

\begin{document}

\preprint{APS/123-QED}

\title{Ferromagnetic quantum critical fluctuations in YbRh$_2$(Si$_{0.95}$Ge$_{0.05}$)$_2$}
\author{P. Gegenwart}
\thanks{also at: School of Physics and Astronomy, University of St. Andrews, North Haugh, St. Andrews KY16 9SS, UK}
\author{J. Custers}
\thanks{present address: Institute for Solid State Physics, University of Tokyo, Chiba 277-8581, Japan}
\author{Y. Tokiwa}
\author{C. Geibel}
\author{F. Steglich}
\affiliation{Max-Planck Institute for Chemical Physics of Solids,
D-01187 Dresden, Germany}

\date{\today}

\begin{abstract}
The bulk magnetic susceptibility $\chi(T,B)$ of
YbRh$_2$(Si$_{0.95}$Ge$_{0.05}$)$_2$ has been investigated
close to the field-induced quantum critical point at
$B_c=0.027$~T. For $B\leq 0.05$~T a Curie-Weiss law with a negative
Weiss temperature is observed at temperatures below 0.3~K. Outside this region, the susceptibility indicates
ferromagnetic quantum critical fluctuations: $\chi(T)\propto
T^{-0.6}$ above 0.3~K, while at low temperatures the Pauli
susceptibility follows $\chi_0\propto (B-B_c)^{-0.6}$ and scales
with the coefficient of the $T^2$ term in the electrical
resistivity. The Sommerfeld-Wilson ratio is highly enhanced and
increases up to 30 close to the critical field.
\end{abstract}

\pacs{71.10.HF,71.27.+a} \maketitle

Landau's Fermi liquid theory has been successfully used to
describe the low-temperature behavior of strongly correlated
electron systems. Starting from a Fermi gas, this model
introduces the many-body interactions in a phenomenological way.
It is based on the concept of elementary excitations, called
quasiparticles, showing a one-to-one correspondence to the free
electron (or hole) excitations of the Fermi gas. Furthermore, the
quasiparticle motion can be described by a generalized Boltzmann
equation. The quasiparticle excitations thus lead to a linear in
temperature ($T$) specific heat, $C=\gamma_0 T$, and a constant
Pauli susceptibility $\chi_0$ at low temperatures as well as a
temperature independent rate $\propto A$ of
quasiparticle-quasiparticle collisions causing an electrical
resistivity contribution $\Delta\rho=AT^2$. The electronic
correlations result in a renormalization of the effective mass of
the quasiparticles which in case of the heavy fermion (HF) systems
can exceed the bare electron mass by a factor up to 1000. This
causes huge values of $\gamma_0$, $\chi_0$ and $A$ that roughly
scale like $\gamma_0\propto \chi_0\propto \sqrt{A}$. Recently,
much interest has been focused on how the properties of the heavy
Landau Fermi liquid (LFL) state evolve if these materials are
tuned into a long-range magnetically ordered state \cite{Stewart}. The important question arises whether the heavy
quasiparticles retain their itinerant character and form a
spin-density wave (SDW) at the quantum critical point (QCP) or, alternatively, decompose
due to the destruction of the Kondo screening. In the latter case
the magnetic order is caused by localized $f$-electrons that do
not contribute to the Fermi surface \cite{Coleman}. In order to
address this question, detailed experiments on the nature of the
quantum critical state in the two prototypical materials
CeCu$_{5.9}$Au$_{0.1}$ \cite{Loehneysen} and YbRh$_2$Si$_2$
\cite{Trovarelli} have been performed.

In CeCu$_{5.9}$Au$_{0.1}$ the static susceptibility has been found
to obey a modified Curie-Weiss (CW) law \cite{Schroeder}
\begin{equation}
\chi^{-1}({\bf q},T)=(T^\alpha+(-\Theta({\bf q}))^\alpha)/c
\label{modCW}
\end{equation}
with a fractional exponent $\alpha\simeq 0.75$. The Weiss
temperature $\Theta({\bf q})<0$ is a function of ${\bf q}$ and
vanishes at the critical wave vector ${\bf q=Q}$ of the nearby
antiferromagnetic (AF) order \cite{Schroeder}. Furthermore, the
dynamical susceptibility follows energy over temperature scaling
and the bulk (${\bf q}=0$) susceptibility obeys magnetic field
over temperature scaling, both with the same fractional exponent
$\alpha$ obtained from the modified CW law \cite{Schroeder}. The
momentum independence in the critical response observed in these
experiments led to the proposal of a {\it locally critical}
scenario for the HF QCP \cite{Si}.

YbRh$_2$Si$_2$ is a clean and stoichiometric HF system located
extremely close to the border of long-range magnetic order and
shows pronounced non-Fermi liquid behavior in thermodynamic,
electrical transport and magnetic properties
\cite{Custers,Trovarelli,Gegenwart,Kuechler,Ishida,Gegenwart
SCES04}. Very weak AF ordering at $T_N=70$~mK can be driven to
zero by a small critical magnetic field $B_c$ of 0.06~T applied
in the easy magnetic plane perpendicular to the crystallographic
$c$-axis \cite{Gegenwart}. In
YbRh$_2$(Si$_{0.95}$Ge$_{0.05}$)$_2$ the partial substitution of
Si-atoms with the larger but isoelectronic Ge reduces $T_N$ and
$B_c$ far closer towards zero (20~mK and 0.027~T, see Figure 4).
The observed divergences of both the quasiparticle mass and
Gr\"uneisen ratio \cite{Custers,Kuechler} exclude the SDW
description of the QCP in this system. Temperature over magnetic
field scaling in thermodynamic and transport properties indicates
that the characteristic energy of the heavy quasiparticles is
governed only by the ratio of the thermal energy to the magnetic
field difference $b=B-B_c$ and vanishes at $b\rightarrow 0$
\cite{Custers,Gegenwart SCES04}. The observed disparity in the
temperature dependence of the electrical resistivity and specific
heat at $b=0$ suggests a break-up of the heavy quasiparticles in
the approach of the QCP \cite{Custers}. This is
consistent with the observation of the Yb$^{3+}$ electron spin
resonance at temperatures at least down to 2~K, i.e. well below
the single-ion Kondo scale of 25~K in that system, that
highlights the emergence of large unscreened local magnetic
moments close to the QCP \cite{Sichelschmidt}.

In this Letter, we use low-temperature measurements of the bulk magnetic susceptibility $\chi(T,B)$ to investigate
the quantum critical behavior in
YbRh$_2$(Si$_{0.95}$Ge$_{0.05}$)$_2$. Our results highlight that
the quantum critical fluctuations in this system have a very
strong ferromagnetic (FM) component and are thus unique among all
other quantum critical HF systems, including
CeCu$_{5.9}$Au$_{0.1}$.

The measurements were performed on pieces of a high-quality
single crystal of YbRh$_2$(Si$_{0.95}$Ge$_{0.05}$)$_2$ studied
previously by specific heat and electrical resistivity
\cite{Custers}, as well as thermal expansion measurements
\cite{Kuechler}. The residual resistivity of the crystals amounts
to 5 $\mu\Omega$cm.  We obtain the magnetic susceptibility
$\chi(T,B)$ from either low-temperature ac-susceptibility or
dc-magnetization measurements. The ac susceptibility was
determined with a low-frequency (16.67 Hz) field modulation of
0.1 mT. Constant fields $B$ have been superposed to the modulation
field using a superconducting 20~T magnet. A $B=0$ study has already
been published in \cite{Gegenwart Acta}. The dc-magnetization
measurements were performed utilizing a high-resolution Faraday
magnetometer.

\begin{figure}
\centerline{\includegraphics[width=8cm,keepaspectratio]{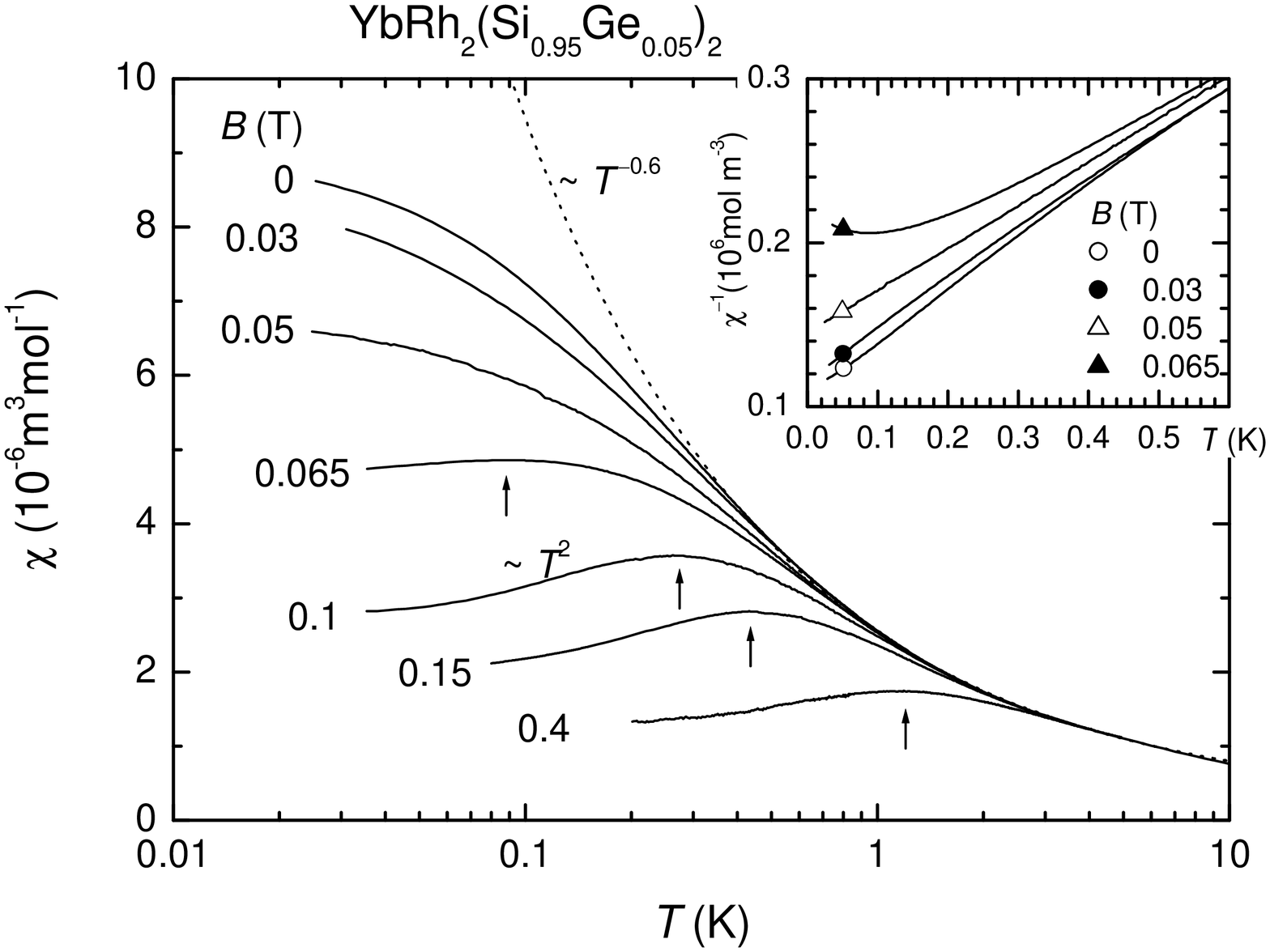}}
\caption{Low-frequency ac susceptibility $\chi$ vs $T$ (on a
logarithmic scale) and $\chi^{-1}$ vs $T$ (inset) of
YbRh$_2$(Si$_{0.95}$Ge$_{0.05}$)$_2$ at varying superposed static
magnetic fields applied perpendicular to the $c$-axis. Dotted
line indicates $(\chi(T)-c)\propto T^{-0.6}$ with $c=0.215\times
10^{-6}$m$^3$mol$^{-1}$. Arrows indicate susceptibility maxima.}
\label{fig1}
\end{figure}

Figure 1 displays the temperature dependence of the magnetic ac
susceptibility of YbRh$_2$(Si$_{0.95}$Ge$_{0.05}$)$_2$ at
different fields $B$, applied in the easy magnetic plane
perpendicular to the $c$-axis. We first concentrate on the $B=0$
data. Upon cooling to below 10~K, a strong increase is observed
that, above 0.3~K, can be approximated by a power law divergence
$\Delta\chi\propto T^{-0.6}$. Here $\Delta\chi$ is the
susceptibility after subtraction of a small temperature
independent contribution that amounts to 2\% of the total
susceptibility at 0.02~K. The previous attempt \cite{Gegenwart Acta}
to fit the data with an exponent of 0.75 is much less satisfactory. At lower temperatures, $\chi(T)$ tends
to saturation and is well described by a CW law with a negative
Weiss temperature of $\Theta=-0.32$~K similar to that
found for pure YbRh$_2$Si$_2$ at $T_N<T\leq 0.3$~K
\cite{Custers}. The value of the slope in $\chi^{-1}(T)$
indicates a large effective moment $\mu_{eff}\approx 1.4\mu_B$
per Yb$^{3+}$ and the sign of the Weiss temperature suggests some
AF correlations \cite{Custers}. Note that the temperature
dependence both above and below 0.3~K is different to that found
in the bulk susceptibility of CeCu$_{5.9}$Au$_{0.1}$, cf. Eq.
(\ref{modCW}). No signature of magnetic ordering is observed
because the experiments have been performed above 20~mK. Upon
superposing constant fields $B$ to the field modulation, the
low-temperature susceptibility decreases. For small fields the
temperature dependence does not change significantly and the CW
law is observed for $B\leq 0.05$~T (see inset). At fields larger
than 0.05~T, the behavior changes drastically: Upon cooling,
$\chi(T)$ passes through a maximum followed by a $T^2$ dependence
at low temperatures, indicating the formation of a field-induced
LFL state \cite{mag} also observed in specific heat and
electrical resistivity measurements \cite{Custers}. The
extrapolated saturation values $\chi_0(B)$ therefore represent
the Pauli susceptibility.

\begin{figure}
\centerline{\includegraphics[width=8cm,keepaspectratio]{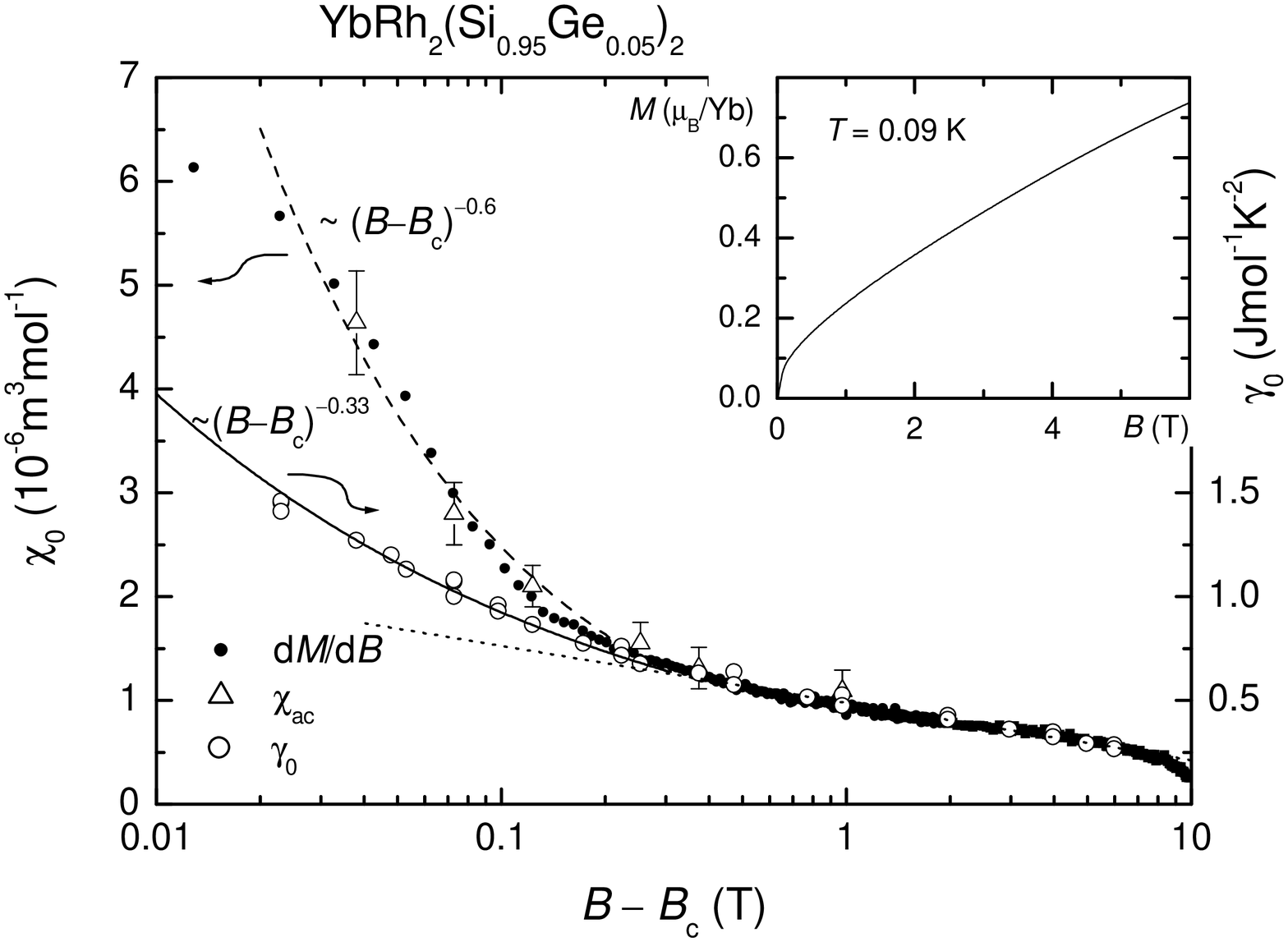}}
\caption{Field dependence of the Pauli magnetic susceptibility
$\chi_0$ determined from the differential susceptibility $dM/dB$
at 0.09~K (solid circles, left axis) and $T\rightarrow 0$
extrapolation of the ac susceptibility $\chi(T)$ (open triangles,
left axis) as well as specific heat coefficient $\gamma_0$
(\cite{Custers}, open circles, right axis). Solid, dashed and
dotted lines indicate $\gamma_0\propto(B-B_c)^{-0.33}$
($B_c=0.027$~T), $\chi_0\propto(B-B_c)^{-0.6}$ and logarithmic
behavior, respectively. Inset shows magnetization $M(B)$ at
$T=0.09$~K.} \label{fig2}
\end{figure}

Next, we focus on the field-dependence of $\chi_0(B)$ in the
approach of the QCP at $B_c=0.027$~T. In Figure 2, we show that
the Pauli susceptibility, determined as discussed above from the
saturation values of isofield ac-susceptibility measurements
(open triangles), agrees well with the slope $dM(B)/dB|_{T=const}$
(solid circles) of the low-temperature dc magnetization (see
inset). The specific heat coefficient in the field-induced LFL
state at $B>B_c$ has been found to diverge in the approach of the
critical field \cite{Custers} and we now compare its field
dependence with that of the Pauli-susceptibility. For fields
larger than about 0.3~T, both properties show a very similar field
dependence (cf. Figure 2). Below 0.3~T, they deviate from each
other, both showing a stronger than logarithmic increase. Whereas
$\gamma(b)\propto b^{1/3}$ with $b$ the difference between the
applied and the critical field, $b=B-B_c$ \cite{Custers}, the
Pauli susceptibility can be described by $\chi_0(b)\propto
b^{-0.6\pm 0.1}$. Note, however, that this power-law divergence,
in contrast to that observed for the specific heat coefficient,
does not continue towards $b\rightarrow 0$: The CW law observed
for fields below 0.05~T with a negative Weiss temperature that
does not vanish at the critical field indicates that
$\chi(T\rightarrow 0)$ remains finite at the QCP.

\begin{figure}
\centerline{\includegraphics[width=8cm,keepaspectratio]{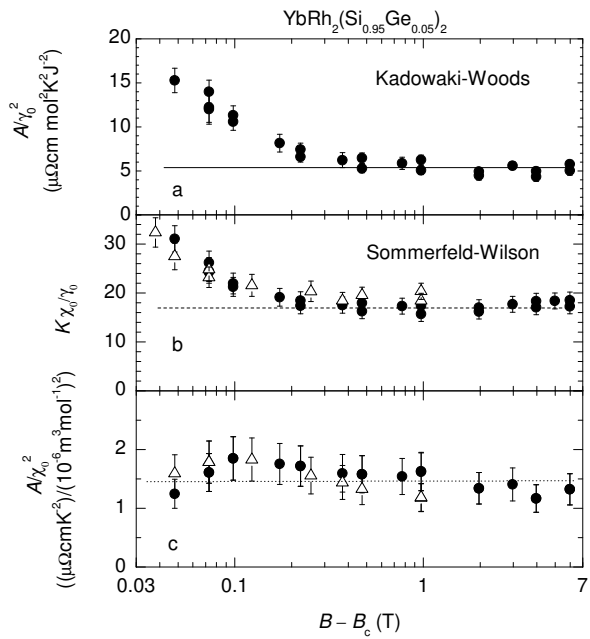}}
\caption{Field dependence of the Kadowaki-Woods ratio
$A/\gamma_0^2$ (a), the Sommerfeld-Wilson ratio
$R_W=K\chi_0/\gamma_0$ 
(b) and the ratio $A/\chi_0^2$
(c) of YbRh$_2$(Si$_{0.95}$Ge$_{0.05}$)$_2$, using $\gamma_0(B)$
and $A(B)$ from Ref. \cite{Custers}. Solid (open) symbols in (b)
and (c) indicate data calculated by using the Pauli susceptibility
$\chi_0$ obtained from the differential susceptibility $dM/dB$ at
90~mK and the $T\rightarrow 0$ extrapolation of the ac
susceptibility (cf. Fig. 2), respectively. Solid, dashed and
dotted lines indicate $A/\gamma_0^2=5.3$
$\mu\Omega$cm~mol$^2$K$^2$J$^{-2}$, $R_W=17.5$ and
$A/\chi_0^2=1.45\times10^{12}
\mu\Omega$cmK$^{-2}/$(m$^3/$mol)$^2$, respectively.} \label{fig3}
\end{figure}

Having determined the field dependence of the Pauli
susceptibility, we may now compare the evolution of the three
characteristic parameters $\chi_0$, $\gamma_0$ and $A$ (the
coefficient of the $T^2$ term in the electrical resistivity) of
the LFL induced for $b>0$ upon tuning the system into the QCP.
This provides information on how the heavy quasiparticles decay
into the quantum critical state. Figure 3a shows the field
dependence of the Kadowaki-Woods ratio \cite{Custers}. At larger
distances from the QCP, $A/\gamma_0^2=const$ is observed.
The weak divergence for $b\rightarrow 0$ indicates that the
characteristic length scale for singular scattering grows much
slower than expected by the itinerant spinfluctuation theory
\cite{Custers}.

Next we focus on the Sommerfeld-Wilson ratio
$R_W=K\chi_0/\gamma_0$, where $K=\pi^2k_B^2/(\mu_0\mu_{eff}^2)$ is a scaling factor which gives
a dimensionless value of $R_W=1$ for the free electron gas.
Whereas electron-phonon interactions enhance
$\gamma_0$ but not $\chi_0$, leading to a reduction of $R_W$, an
enhancement of $R_W$ could be caused by electronic spin-spin
interactions. For Kondo systems, a Sommerfeld-Wilson ratio of 2
is expected \cite{Hewson} as observed in many HF systems
\cite{Fisk}.
Nearly FM metals, due to Stoner enhancement, show very large
values, e.g. $R_W=6-8$ (Pd), 12 (TiBe$_2$), 40 (Ni$_3$Ga)
\cite{Julian} and 10 (Sr$_3$Ru$_2$O$_7$ \cite{Ikeda}). For
YbRh$_2$(Si$_{0.95}$Ge$_{0.05}$)$_2$, as shown in Figure 3b, the
Sommerfeld-Wilson ratio is $b$-independent in the same field
range for which a constant Kadowaki-Woods ratio has been found.
The value of $R_W=17.5\pm 2$ is highly enhanced compared to all
other HF systems. Upon lowering the magnetic field deviation from
the QCP, $R_W$ even increases, reaching a value larger than 30 at
0.065~T, which is the lowest field at which $\chi_0$ could be
determined (see above). This dramatic increase of $R_W$
highlights the importance of FM fluctuations in the approach of
the QCP.

In Figure 3c, the field dependence of the ratio $A/\chi_0^2$,
which compares the quasiparticle-quasiparticle scattering cross
section with the Pauli susceptibility, is shown. In contrast to
both the Kadowaki-Woods and Sommerfeld-Wilson ratio, $A/\chi_0^2$
is approximately constant in the entire field interval above
0.065~T. Since the electrical resistivity is strongest influenced
by large-{\bf q} scattering \cite{Resistivity}, one would not
expect the $A$-coefficient to scale with the ${\bf q}=0$
susceptibility in the approach of an {\it antiferromagnetic} QCP.
The fact that $A/\chi_0^2\approx const$ over more than two
decades in the field-deviation from the QCP thus provides
evidence for FM (${\bf q}=0$) quantum critical fluctuations in
YbRh$_2$(Si$_{0.95}$Ge$_{0.05}$)$_2$.

\begin{figure}
\centerline{\includegraphics[width=7.5cm,keepaspectratio]{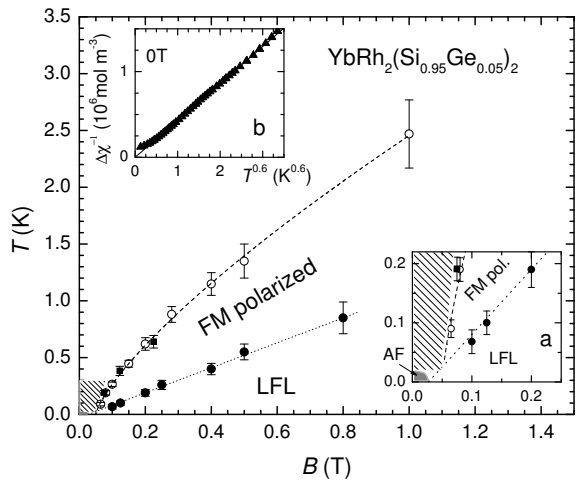}}
\caption{$T$-$B$ phase diagram for
YbRh$_2$(Si$_{0.95}$Ge$_{0.05}$)$_2$, $B\perp c$. Open and closed
circles indicate temperatures of maxima in $\chi(T)$ (cf. arrows
in Fig.~1) and $C(T)/T$ \cite{Custers} at various different
fields, respectively. Positions of $\chi(T)$ maxima for
YbRh$_2$Si$_2$ \cite{Trovarelli} are indicated by small solid
squares. Dashed and dotted lines represent
$2.55$~KT$^{-0.75}\times(B-0.05$~T$)^{0.75}$ and
$1.1$~KT$^{-1}\times(B-0.027$~T$)$, respectively. AF state marked
by grey solid region very close to the origin. Slanted lines
indicate regime where $\chi(T)$ follows Curie-Weiss law. Labels see text. Inset a enlarges region close to origin. Inset b
displays susceptibility increment as $\Delta\chi^{-1}$ vs
$T^{0.6}$ with $\Delta\chi=\chi(T)-0.215\times
10^{-6}$~m$^3$mol$^{-1}$. Solid line indicates
$\Delta\chi^{-1}\propto T^{0.6}$.}\label{fig4}
\end{figure}

Figure 4 shows the temperature-field diagram for
YbRh$_2$(Si$_{0.95}$Ge$_{0.05}$)$_2$ including regimes of
different magnetic response. The AF state close to the origin is
surrounded by a regime below 0.3~K that extends to fields up to
0.05~T (shaded area), in which the susceptibility follows a CW
law with a negative Weiss temperature, indicating predominant AF
correlations. Outside this region, the quantum critical behavior
is dominated by FM fluctuations: i) $\chi_0(b)$ follows a
$b^{-0.6}$ dependence and ii) the temperature dependent part,
$\Delta\chi(T)$, diverges as $T^{-0.6}$ for $T>0.3$~K suggesting a
divergent ${\bf q}=0$ susceptibility (see also inset of Figure
4). A similar temperature dependence has been observed in the
$^{29}$Si NMR-derived Knight shift $K_s(T,B)$ of YbRh$_2$Si$_2$
\cite{Ishida}. In these experiments, outside a narrow region
close to the critical field, the Korringa ratio $(1/T_1T)/K_s^2$
with $K_s$ and $1/(T_1T)$, being proportional to the bulk
susceptibility and the {\bf q}-averaged dynamical spin
susceptibility, respectively, is constant, with a value similar as
found for nearly ferromagnetic metals \cite{Ishida}. This
suggests that the inverse of the zero-field susceptibility,
plotted versus $T^{0.6}$ in the inset of Fig.~4, is effectively
{\bf q}-independent above 0.3~K. Such behavior is even "more
local" than that described in the {\it locally-critical} scenario
\cite{Si} and very different to the case of
CeCu$_{5.9}$Au$_{0.1}$ (cf. Eq. \ref{modCW}) for which latter
system the Weiss temperature is strongly {\bf q} dependent and
vanishes only for ${\bf q=Q}$, i.e. at the critical
antiferromagnetic wave vector \cite{Schroeder}.

Finally, we discuss the characteristic maximum in $\chi(T)$ whose
position shifts to lower temperatures with decreasing magnetic
field extrapolating towards $B^\star\approx0.05$~T (see dashed
line in Figure 4). Very similar behavior is observed for pure
YbRh$_2$Si$_2$ as well \cite{Trovarelli}, the positions of the
susceptibility maxima being not affected by the Ge-substitution
(cf. open circles and solid squares in Figure 4). These positions
of susceptibility maxima define a line in the
temperature-field plane along which the magnetization slope
$dM/dB$ is most sensitive to a change of the applied magnetic
field. This characteristic field increases with temperature as
expected for a FM polarization of fluctuating magnetic moments.
Most interestingly, the {\it ferromagnetic} fluctuations are
unaffected by the Ge-substitution in
YbRh$_2$(Si$_{0.95}$Ge$_{0.05}$)$_2$ that has strong influence on
the {\it antiferromagnetic} order, leading to a roughly threefold
reduction of the ordering temperature, ordered moment
\cite{Gegenwart SCES04} and critical magnetic field compared to
the parent compound YbRh$_2$Si$_2$. This suggests the FM
fluctuations to be not directly correlated to the AF-QCP at
$B_c=0.027$~T in the Ge-substituted system. Indeed, the line of
susceptibility maxima is different from the cross-over line
determined from the maxima in the specific heat coefficient
$C(T)/T$ that terminates at $B=B_c=0.027$~T, i.e. $\approx
{1\over 2} B^\star$ (\cite{Custers}, see solid circles in Figure
4).

To summarize, the strong increase of the bulk susceptibility
towards low temperature, the highly enhanced Sommerfeld-Wilson
ratio and the field-independence of $A/\chi_0^2$ indicates that
YbRh$_2$(Si$_{0.95}$Ge$_{0.05}$)$_2$ is located very close to a FM
instability. Recent experiments on itinerant ferromagnets have
revealed a first-order instead of a continuous suppression of the
ordering \cite{Uhlarz ZrZn2}. It
has also been argued that close to a FM QCP a nonanalytic term in
the free energy generates first-order behavior \cite{Belitz}. In
$4f$-based heavy fermion systems, no evidence for a FM QCP has yet
been found, instead these systems first undergo a transition to an
AF state before getting paramagnetic \cite{Sullow,Eichler}.
YbRh$_2$(Si$_{0.95}$Ge$_{0.05}$)$_2$ is thus unique as the
quantum critical behavior is dominated by FM fluctuations over
wide ranges of the $T-B$ plane, except for fields close to the
critical field and temperatures below 0.3~K.

Work supported in part by the Fonds der Chemischen Industrie.


\begin{references}
\bibitem{mag} As discussed in \cite{Gegenwart SCES04}, the
temperature dependent part of the susceptibility varies as
$T^2b^{-7/3}$ in the field-induced LFL, obeys $T/b$ scaling
($b=B-B_c$) and is consistent with the field-dependence of the
magnetic entropy.
\end{references}
\end{document}